\documentclass[reprint,aps,prb,amsmath,amssymb]{revtex4-1}

\usepackage[]{graphicx}
\usepackage[]{units}
\usepackage{times}
\usepackage{bm}
\usepackage{ulem}		
\usepackage{color}

\newcommand{\comment}[1]{}

\renewcommand{\emph}{\textit}

\begin{document}

\title{Impact of optically pumped non-equilibrium steady states on luminescence emission of atomically-thin semiconductor excitons}
\author{Malte Selig$^1$}\email{malte.selig@tu-berlin.de}
\author{Dominik Christiansen$^1$}
\author{Manuel Katzer$^1$}
\author{Mariana V. Ballottin$^2$}
\author{Peter C. M. Christianen$^2$}
\author{Andreas Knorr$^1$}
\affiliation{$^1$ Nichtlineare Optik und Quantenelektronik, Institut f\"ur Theoretische Physik, Technische Universit\"at Berlin,  10623 Berlin, Germany}
\affiliation{$^2$ High Field Magnet Laboratory (HFML-EMFL), Radboud University,
Toernooiveld 7, Nijmegen 6525 ED, the Netherlands}

\begin{abstract}

The interplay of the non-equivalent corners in the Brillouin zone of transition metal dichalcogenides have been investigated extensively. While experimental and theoretical works contributed to a detailed understanding of the relaxation of selective optical excitations and the related relaxation rates, only limited microscopic descriptions of stationary experiments are available so far. In this manuscript we present microscopic calculations for the non-equilibrium steady state properties of excitons during continuous wave pumping. We find sharp features in photoluminescence excitation spectra and degree of polarization which result from phonon assisted excitonic transitions dominating over exciton recombination and intervalley exchange coupling.

\end{abstract}

\maketitle



\textit{Introduction:} Tightly bound excitons and strong exciton interactions in transition metal dichalcogenides have stimulated research over the last years\cite{Wang2018,varsano2020monolayer,gies2021atomically,dong2021direct}. Optically addressable excitons are located at the non-equivalent $K$/$K'$ valleys in the hexagonal Brillouin zone, which can be addressed with light of opposite circular polarization. The valley lifetime of selectively excited valley excitons, is typically measured via pump probe experiments\cite{Wang2013,kumar2014valley,Conte2015,Schmidt2016,Wang2018,ZWang2018}, time resolved luminescence experiments\cite{yan2015valley}, or stationary luminescence\cite{Wang2015,kim2017electrical,Smolenski2016,yan2015valley,zeng2012valley,kioseoglou2012valley,Tornatzky2018,jadczak2019room}. Intervalley exchange coupling has been shown to be the major source of relaxation between the valleys after optical excitation\cite{Yu2014,Glazov2014,selig2019ultrafast}. However as revealed recently, in the presence of energetically low lying momentum-indirect exciton states\cite{Qiu2015,Selig2018} the intervalley exchange coupling is efficiently suppressed\cite{selig2020suppression} such that other spin relaxation mechanisms as Dyakonov-Perel\cite{Wang2014,Wang2014b} or Elliott Yafet\cite{Wang2014b,molina2017ab,ZWang2018} gain importance. Energetically low lying momentum-indirect states, such as $( K , \Lambda )$ excitons in WSe$_2$, are present in many TMDC materials. In contrast, MoSe$_2$ is considered as a candidate where momentum indirect, i.e. dark, states are of minor importance and intervalley relaxation is governed through exchange coupling. While the temporal behavior of optical excitations including the calculation of the different scatterings is well understood in MoSe$_2$ and also other materials, so far no theoretical studies focused on the understanding of stationary experiments. As one candidate, stationary photoluminescence excitation (PLE) experiments are carried out to study the relaxation pathways of optical excitations. Here, several studies reported on the emergence of phonon-replica which appear at excitation energies above the A exciton transition\cite{chow2017phonon,chow2017unusual,shree2018observation}.

In this manuscript we present a microscopic investigation of the interplay of exchange coupling and exciton phonon interaction in monolayer MoSe$_2$ in stationary experiments under circular polarized continuous wave (cw) laser excitation, this way analyzing a prototype of atomically-thin semiconductors with a bright ground state. We take the detuning $\Delta$ of the laser frequency located above the $A$ exciton transition into account and explicitly evaluate its influence on the photoluminescence excitation spectrum (PLE) and the degree of polarization (DoP) of the emitted light. Both, the PLE and the DoP as a function of the detuning exhibit distinct features which can be traced back to phonon assisted excitonic transitions. Depending on the range of the detuning $\Delta$, we identify different microscopic processes which lead to the formation of specific spectral PLE features: (i) for increasing, but small, detuning of few meV, the DoP decreases due to increasing off-resonance. At larger detunings above \unit[30]{meV}, (ii) phonon assisted transitions lead to an enhanced degree of polarization, namely due to intravalley optical phonon emission processes: (iii) two $\Lambda / M$ phonon emission processes, being assisted via virtual transitions at high energy excitonic valleys and (iv) the stepwise relaxation due to two intravalley optical phonon emission processes. The calculated peaks in the PLE agree well with recently reported measurements\cite{chow2017phonon}. Moreover, we calculate the effective exciton temperatures and reveal that cw excitations lead to significant heating of excitons such that they can hardly be assumed to be in equilibrium with the lattice in many experimental situations.

\begin{figure*}[t!]
 \begin{center}
\includegraphics[width=1.0\linewidth]{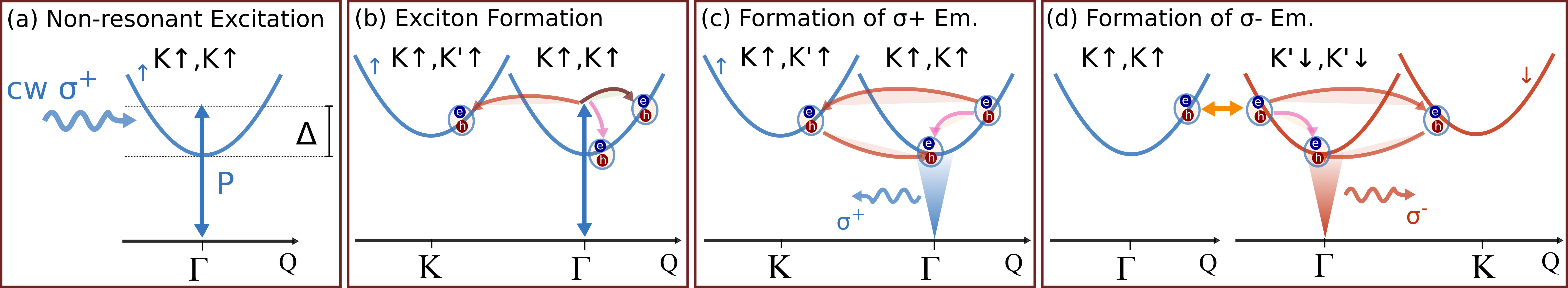}
 \end{center}
 \caption{\textbf{Schematic illustration of the competing processes.} (a) An excitonic coherence is non-resonantly pumped by the continuous wave light. (b) Exciton phonon scattering mediates the formation of an incoherent occupation of exciton states. Here, optical $\Gamma$ phonon scattering (pink arrow), acoustic $\Gamma$ phonon scattering (brown arrow) and intervalley phonon scattering (red arrow) contributes. (c) Exciton phonon scattering mediates the thermalization and population of states within the radiative cone (blue shaded area) (d) For the population of the $\sigma_-$ radiative cone (red shaded area) a stepwise process is required: (i) Intervalley exchange coupling (yellow arrow) has to generate $\downarrow$ excitons and subsequent exciton phonon scattering populates the radiative cone.}
 \label{schema}
\end{figure*}

All in all, as the microscopic origin for this observation we identify the complex interplay of hot exciton formation via non-resonant cw excitation, intervalley exchange coupling, exciton-phonon scattering and decay of excitons which contribute to the formation of a non-equilibrium steady state (NESS). The interplay of the individual contributions determining the PLE are illustrated in detail in figure \ref{schema}. First, cf. figure \ref{schema} (a), the excitonic coherence $P$ is excited with a $\sigma_+$ polarized cw light pulse with a certain detuning $\Delta$ from the $A$ exciton transition in the $( K\uparrow , K\uparrow )$ valley. The excitonic coherence decays via radiative decay\cite{Selig2016,wang2016radiative}, but more importantly through exciton phonon interaction\cite{Selig2016,lengers2020theory}, cf. figure \ref{schema} (b). The latter accounts for the formation of a hot incoherent exciton occupation in direct, as well as momentum-indirect exciton states. In the course of the formation of photoluminescence, two competing processes occur: the direct relaxation of excitons due to exciton phonon scattering into the light cone of the pumped valley where $\sigma_+$ polarized photons are emitted, cf. figure \ref{schema} (c), or the subsequent exchange coupling and phonon mediated relaxation to the light cone of the unpumped valley where $\sigma_-$ photons are emitted, cf. figure \ref{schema} (d). The different efficiencies of the latter relaxation mechanisms determines the ratio of the intensity of $\sigma_+$ and $\sigma_-$ polarized light. While all processes appear simultaneously, the resulting exciton distribution is in a non-equilibrium steady state (NESS), which in general depends on the phonon temperature, the detuning of the exciting light field as well as on the general lifetime of the excitons.

\textit{Theory:} In order to study the non-equilibrium steady states of excitons, i.e. the interplay of cw excitation, phonon-mediated thermalization, exchange-driven intervalley scattering and exciton recombination, in MoSe$_2$ during cw excitation, we start with the parametrization of the excitonic Hamiltonian\cite{Katsch2018} building on DFT calculations\cite{Kormanyos2015,Li2014,Trolle2017}. The Hamiltonian as well as all input parameters are summarized in the reference \onlinecite{selig2020suppression} in the tables IV to IX (open access). The next step is to calculate the equations of motion for the excitonic coherence $P^{\xi_h \xi_e}(t)$ and incoherent exciton occupation $N^{\xi_h \xi_e}_\mathbf{K}(t)$ with the compound valley spin of holes and electrons $\xi_{h/e}=(i_{h/e},s_{ h/e})$ with valley $i_{h/e}$ and spin $s_{h/e}$ of the carriers and Fourier component of the center of mass motion $\mathbf{K}$. We take into consideration the exciton-light, exciton-phonon and intervalley exchange coupling such that we end at the Bloch equation for the excitonic coherence and the Boltzmann equation for incoherent exciton occupations\cite{Selig2019}, cf. supplementary material section I. 
While our microscopic evaluation reveals a radiative lifetime of excitons in the order of hundreds of ps\cite{Selig2018}, we include an additional decay $\tau_{dark}^{-1}$ to our equations of motion, which can be attributed to dark recombination processes, e.g. exciton-exciton annihilation\cite{erkensten2021dark,steinhoff2021microscopic}. For our investigation we excite the excitonic coherence with a cw light field which is detuned from the excitonic energy by $\Delta$. We explicitely evaluate the steady state of the exciton occupation and calculate the polarization-resolved stationary photoluminescence with polarization $\sigma = \sigma_+,\sigma_-$\cite{Kira1999,Thranhardt2000}

\begin{align}
I^{\sigma}(\Delta)\propto\frac{2 \pi}{\hbar}\hspace{-2pt}\sum_{{\mathbf{K},K_z},\xi} |d^{\xi \sigma}_{\mathbf{K}}|^2  N_\mathbf{K}^{\xi\xi}(\Delta,t\rightarrow \infty) \delta( \Delta E^{\xi \sigma}_{\mathbf{K},K_z} ),\label{eqPL}
\end{align}
with the dipole momentum $d^{\xi \sigma}_{\mathbf{K}}$. The appearing delta function ensures the energy and momentum conservation during the photoemission. The emission depends on the amout of excitons $N_\mathbf{K}^{\xi\xi}$ which is located in the radiative window (determined via the delta function), $\Delta E_{\mathbf{K},K_z}^{\xi\sigma} = E_\mathbf{K}^{\xi\xi} - \hbar \Omega_{\mathbf{K},K_z}^\sigma$, with the exciton dispersion $E_\mathbf{K}^{\xi\xi}$ and the photon dispersion $\hbar \Omega_{\mathbf{K},K_z}^\sigma$ with the three dimensional wave vector $(\mathbf{K},K_z)$. Note, that the stationary luminescence depends via the occupation $N_\mathbf{K}^{\xi\xi}$ on the laser detuning via the exciton formation process, supplementary material in section I, such that we write this dependence explicitly.

For the following, we numerically evaluate the exciton dynamics and the polarization resolved photoluminescence for the material MoSe$_2$ on a quartz substrate. However we have carefully checked that the described effects do not change qualitatively if we use different dielectric constants of the environment. If not stated differently we use a lattice temperature of \unit[7]{K} and an exciton lifetime $\tau_{dark}$ of \unit[5]{ps}. We take exciton phonon coupling with the optical $TO$ and $A'$ and the acoustic $LA$ and $TA$ modes into account, cf. the detailed discussion in the supplementary material, section I.
\begin{figure}[h!]
 \begin{center}
\includegraphics[width=1.0\linewidth]{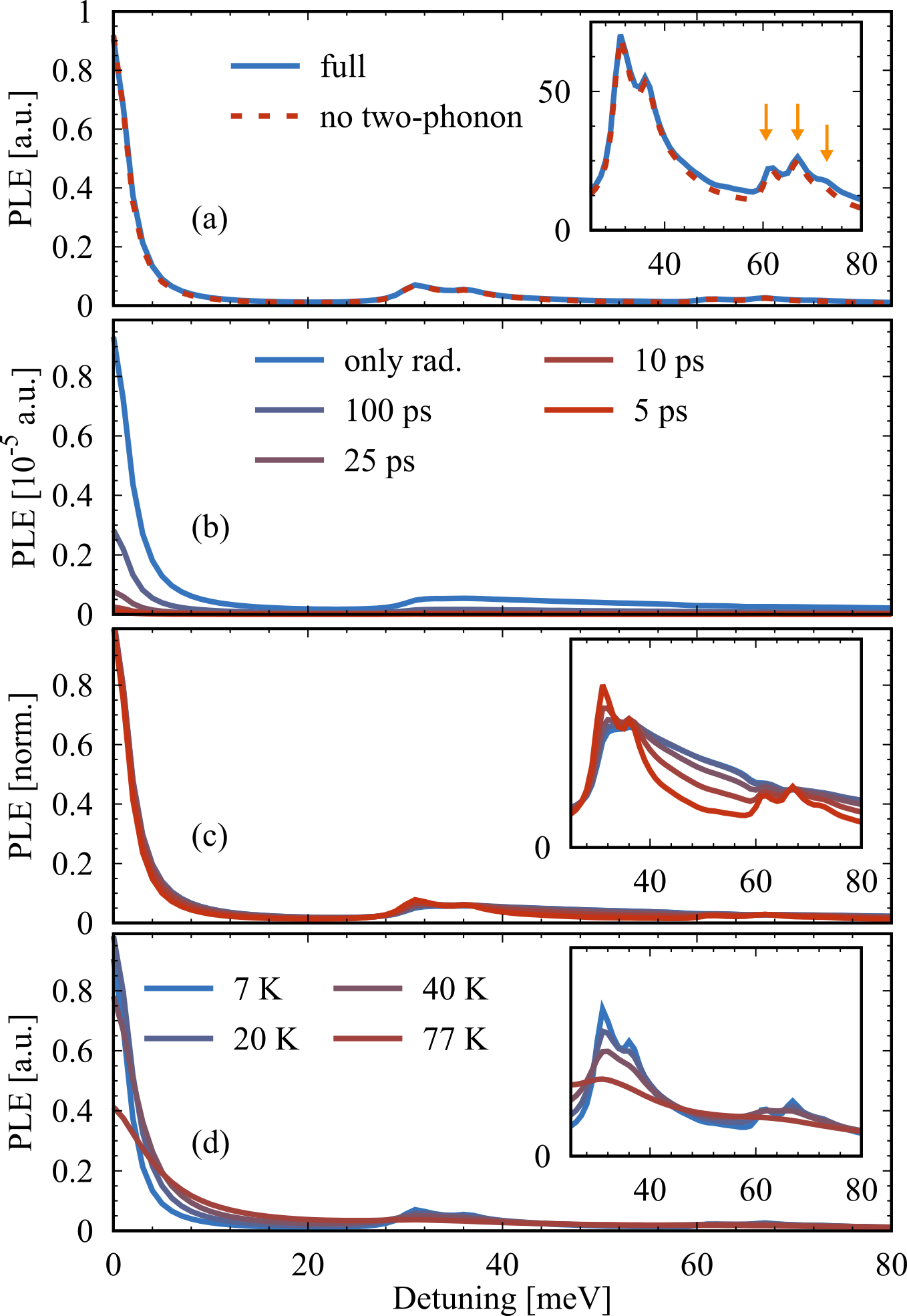}
 \end{center}
 \caption{\textbf{Photoluminescence Excitation Spectrum (PLE)} at \unit[7]{K}. (a) Contributions to the PLE for an exciton lifetime of \unit[5]{ps}. The inset shows the spectral region between \unit[25]{meV} and \unit[80]{meV}. (b) for exciton lifetimes of \unit[5]{ps}, \unit[10]{ps}, \unit[25]{ps}, \unit[100]{ps} and only the radiative lifetime. (c) As in figure (b), but normalized to 1. (d) for an exciton lifetime of \unit[5]{ps} at different lattice temperatures of \unit[7]{K}, \unit[20]{K}, \unit[40]{K} and \unit[77]{K}.}
 \label{PLE}
\end{figure}

\textit{Photoluminescence Excitation Spectrum:} Before studying the polarization resolved spectra, we start our investigation with the photolumiscence excitation spectrum (PLE) in the steady state as a function of the laser detuning $\Delta$
\begin{equation}
PLE (\Delta) = \left( I^{\sigma_+} (\Delta) + I^{\sigma_-} (\Delta) \right).
\end{equation}
For long integration times in the experiment (more than nanoseconds), this measure is equivalent to the time integrated photoluminescence intensity.

The PLE is a measure, how many excitations are generated after excitation with a particular laser detuning from the $A$ transition $\Delta$. In figure \ref{PLE} we illustrate an exemplary PLE spectrum as a function of the laser detuning $\Delta$. First of all, we find a general decrease of the PLE due to the detuned excitation of the excitonic transition, cf. supplementary section I. However for detunings larger than \unit[26]{meV}, we find a sharp increase of the PLE which can be traced back to efficient emission via excitonic relaxation pathways which are assisted by the emission of optical intravalley $\Gamma$ and optical and acoustic intervalley phonons. We find sharp features at \unit[30]{meV} and \unit[36]{meV} which originate from exciton-phonon scattering with the optical $\Gamma A'$ and $\Gamma TO$ mode\cite{Li2014}, cf. supplementary figure S3 (b). Similarly, at energies above \unit[60]{meV}, sharp features are observed due to the step wise relaxation of excitons under the emission of two optical $\Gamma$ phonons with exciton states in the $( K\uparrow , K\uparrow) $ and $( K'\downarrow , K'\downarrow) $ valleys as intermediate states. This results in three PLE peaks: the first one involves two $\Gamma A'$ emission processes, the second one involves one $\Gamma A'$ and one $\Gamma TO$ emission processes and third one involves two $\Gamma TO$ emission processes. The intermediate spectral region is structureless. Here, excitons are formed through emission of one optical $\Gamma$ and the subsequent exciton relaxation is driven from acoustic $\Gamma TA$ and $\Gamma LA$ phonon scattering into the radiative cone. Since acoustic phonons have a broad density of states, the corresponding feature in the PLE does not show any significant structure.

Additionally, (cp. dashed and solid line in fig. \ref{PLE}(a)) we find a slight impact of two-phonon processes where two acoustic $M$ phonons are emitted with \textit{virtual} exciton states at the $( K\uparrow , \Lambda'\uparrow) $ and $( K'\downarrow , \Lambda\downarrow) $ as intermediate states. We find a rather broad band of the two phonon processes via virtual states, since different pathways contributes: In MoSe$_2$ there are two acoustic $M$ phonons with energies of \unit[16.4]{meV} (TA) and \unit[19.7]{meV} (LA) and three optical branches with energies of \unit[35.8]{meV} (TO), \unit[37.9]{meV} (LO)  and \unit[27.3]{meV} (A'). Each combination of the latter phonons is involved in the $\alpha,\alpha'$ summation in equation 21, supplementary section II. As a result the two phonon band begins at the energy of two TA phonons  (\unit[32.8]{meV}) and ends at the energy of two LO phonons (\unit[75.8]{meV}). We find the most prominent contribution in the region of \unit[50]{meV} to \unit[60]{meV} which originates from the two phonon scattering involving one acoustic and one optical phonon. In this region, two phonon processes are enhanced in comparison to the two-acoustic phonon region (\unit[32.8-39.4]{meV}) for combinatorial reasons. The calculated peaks in the PLE are in good agreement with a recent measurement of the PLE in MoSe$_2$\cite{chow2017phonon}.

In order to understand the PLE spectrum in more detail, in figure \ref{PLE} (b) we illustrate the PLE spectrum for different exciton lifetimes $\tau_{dark}$. It is known that $\tau_{dark}$ lies in the picosecond region depending on the excitation conditions \cite{steinhoff2021microscopic}. We use a range of \unit[5]{ps} to \unit[100]{ps} to illustrate its impact. The figure clearly indicates that the PLE intensity increases as a function of the exciton lifetime, which is obvious since, with increasing lifetime, also the ratio between radiative and non-radiative recombination increases. Consequently, the influence of the non-radiative contribution decreases resulting in a larger emission amplitude. The normalized PLE spectra for the different exciton lifetimes (c) follow a similar trend in the region of small detunings, since this spectral region the PLE is mainly determined via the absorption lineshape which is independent of the exciton lifetime $\tau_{dark}$. However, we find substantial differences between the spectra with different lifetimes in the spectral region above \unit[30]{meV}: while for short lifetimes, the phonon-induced lines appear as sharp features, an increase of the lifetime results in an out-smearing of these lines. This effect originates from the competition of phonon assisted exciton relaxation and the exciton recombination. For short lifetimes, excitons which are generated energetically slightly above an optical phonon transition decay mainly non-radiatively via $\tau_{dark}$. As the excitons lifetime increases, these hot excitons have enough time to relax under emission of acoustic phonons into the radiative cone and contribute to the PLE. Consequently the PLE increases above the optical $\Gamma$ phonon energy.

Figure \ref{PLE} (d) illustrates the PLE spectrum at a fixed exciton lifetime of $\tau_{dark}=$\unit[5]{ps} at selected lattice temperatures of $T_{lattice}=$\unit[7-77]{K}. We find that for increasing lattice temperature, two main effects can be observed: (i) The region at small detunings as well as all peaks which originate from phonon-induced excitonic transitions are significantly broadened. This can be traced back to the enhanced exciton phonon scattering at elevated lattice temperatures, which results in a broadening of the absorption\cite{Selig2016,Selig2018,khatibi2018impact} and thus a broadening of the lines in the PLE. (ii) We find a general decrease of the PLE as a function of the lattice temperature. This originates from the shape of the excitonic steady state distribution: An increasing temperature results in hotter excitons. Consequently less excitons occupy the bright state relevant for the photoluminescence, which reduces the PLE. However, we find that the ratio between phonon induced peaks and PLE at $\Delta=$\unit[0]{meV} increases as a function of temperature due to more efficient exciton - optical phonon scattering at larger temperatures.

\textit{Degree of Polarization:} The degree of polarization (DoP) of the emitted light as a function of laser detuning $\Delta$ is defined as
\begin{equation}                                                                      
DoP (\Delta) = \frac{I^{\sigma_+}(\Delta) - I^{\sigma_-}(\Delta)}{I^{\sigma_+}(\Delta) + I^{\sigma_-}(\Delta)}.
\end{equation}

\begin{figure}[h!]
 \begin{center}
\includegraphics[width=1.0\linewidth]{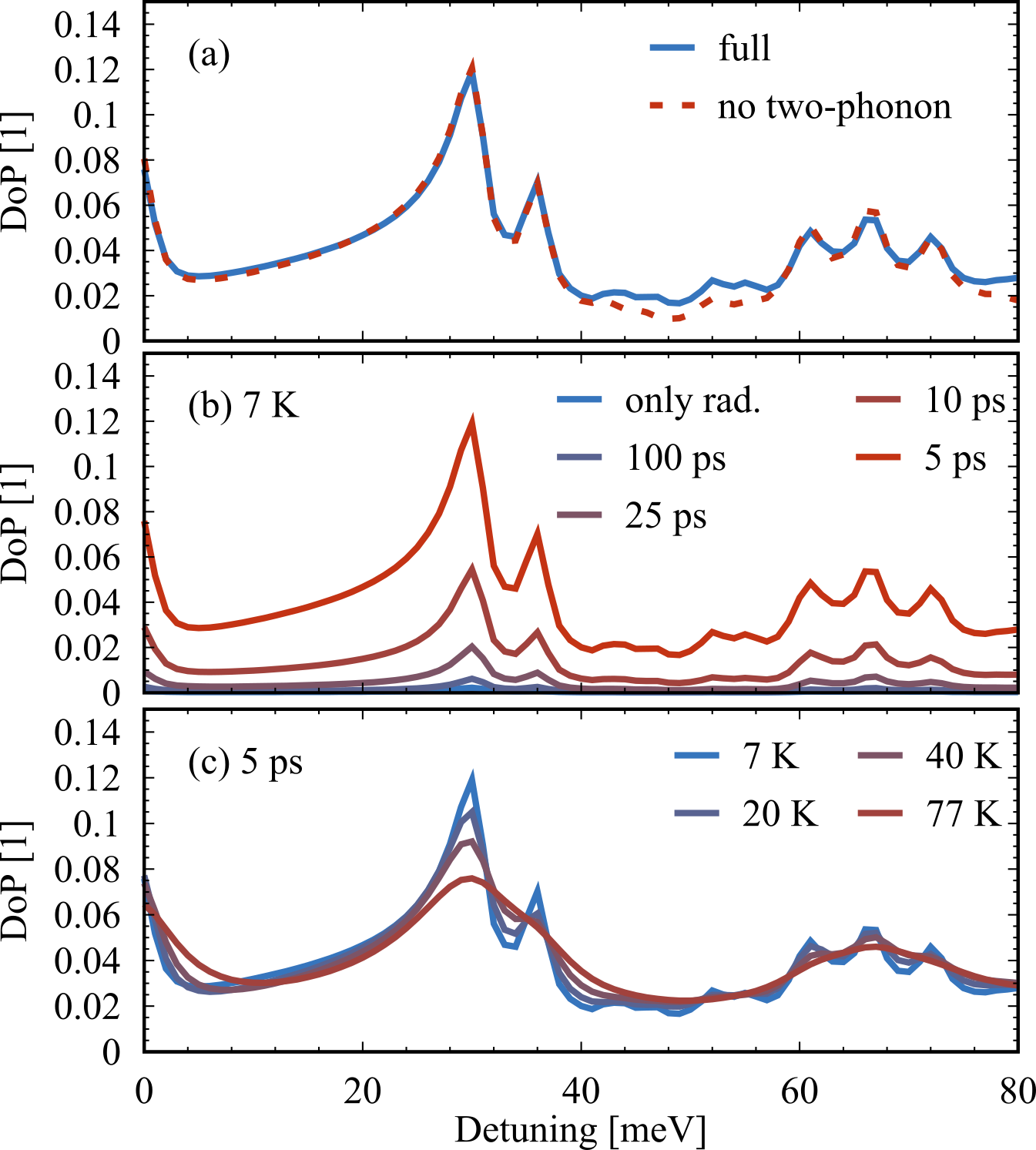}
 \end{center}
 \caption{\textbf{Steady state degree of polarization in MoSe$_2$.} (a) Contributions to the DoP at \unit[7]{K} at an exciton lifetime of \unit[5]{ps}. Studies for the DoP (b) at \unit[7]{K} for different exciton lifetimes and (c) for lattice temperatures $T_{lattice}=$\unit[7-77]{K} for a fixed lifetime of \unit[5]{ps}.}
 \label{Pola}
\end{figure}

Figure \ref{Pola} illustrates the DoP in MoSe$_2$ as a function of laser detuning at a temperature of \unit[7]{K} and an exciton lifetime of $\tau_{dark}=$\unit[5]{ps}. We find several features which originate from the complex interplay of exciton phonon scattering, intervalley exchange coupling, and radiative recombination: (i) for increasing, but small, detuning, we find that the DoP decreases, which originates from the simultaneously increasing exchange coupling element and thus polarization mixing as a function of the center of mass momentum (and kinetic energy) occurs. (ii) At energies above \unit[30]{meV} we find sharp features in the DoP which can be traced back to phonon induced transitions, which favor the relaxation into the light cone in the pumped valley, cf. figure \ref{schema} (c), over the stepwise relaxation into the unpumped valley via exchange and subsequent phonon emission, cf. figure \ref{schema} (d). We identify the same phonon transitions as for the discussion of the PLE, cf. figure \ref{PLE} (a).

To understand the DoP in more detail, in figure \ref{Pola} (b), we show the steady state degree of polarization in MoSe$_2$ at $T_{lattice}=$\unit[7]{K} for different exciton lifetimes $\tau_{dark}$. We find that with decreasing lifetime the degree of polarization increases over the whole spectral range. This can be understood from the fact that the continuous pumping of excitons results in the formation of a non-equilibrium steady state where the overall decay of the excitons significantly influences the steady state distribution. The inclusion of a short exciton lifetime in comparison to the intervalley exchange coupling prevents the excitons from coupling to the other valley relevant for $\sigma_-$ emission resulting in a significant degree of polarization. The observed order of magnitude ranges from less than $0.01$ for only incorporating the radiative decay to approximately $0.12$ for an exciton lifetime of \unit[5]{ps}.

Figure \ref{Pola} (c) illustrates the DoP as a function of laser detuning from the $A$ exciton at an exciton lifetime of \unit[5]{ps} but for varying lattice temperatures $T_{lattice}=$\unit[7-77]{K}. At all temperatures, we find a similar qualitative behavior of the DoP, including a decrease of the DoP at small detunings, and an increased DoP at certain energies which can be related to optical $\Gamma$ phonon induced transitions. Additionally we find a broadening of all involved lines due to more intense exciton phonon interaction.

\textit{Effective Excitonic Temperatures:} After turning on the cw pump, exciton-phonon coupling leads to the formation of a non-equilibrium steady state. Here, one can expect a significant impact of the finite exciton lifetime which will be studied in the current section. To determine the effective exciton temperatures we perform a Boltzmann fitting of the excitonic steady state distributions for different exciton lifetimes $\tau_{dark}$ and laser detuning $\Delta$. The steady state distributions are discussed in the supplementary material, section III.

\begin{figure}[h!]
 \begin{center}
\includegraphics[width=1.0\linewidth]{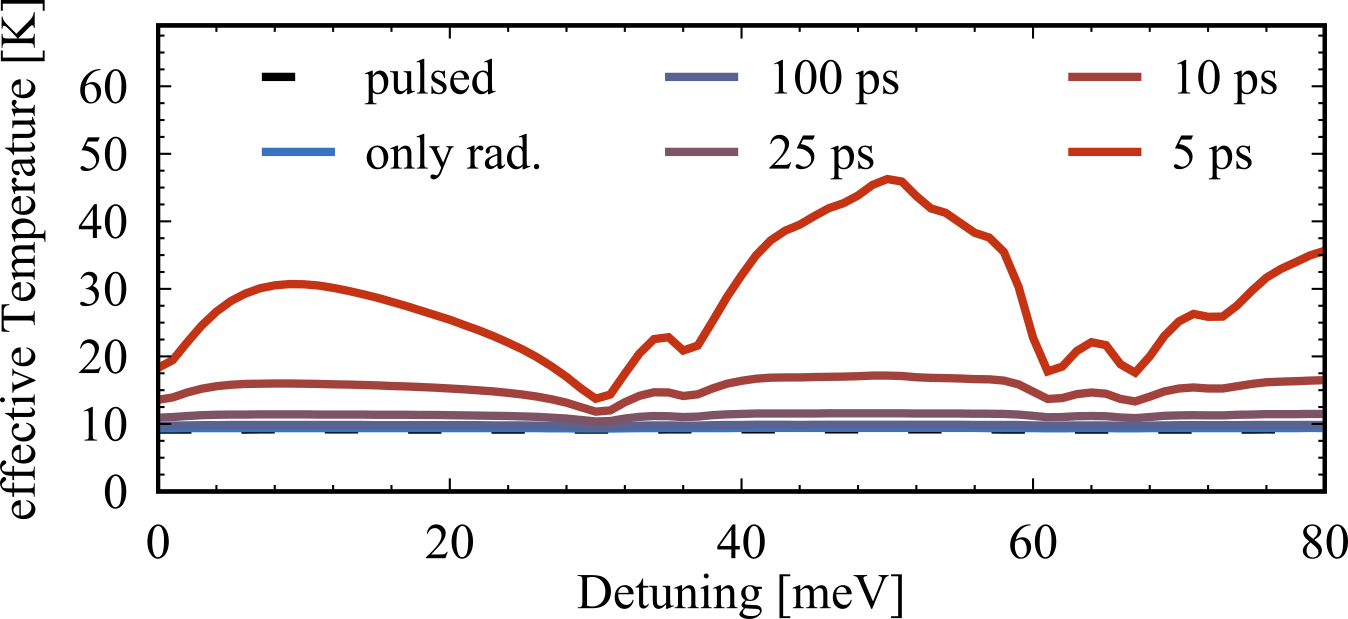}
 \end{center}
 \caption{\textbf{Effective Temperatures.} Extracted effective exciton temperature at a lattice temperature of $T_{lattice}=$\unit[7]{K} in the steady state as a function of the laser detuning for different exciton lifetimes.}
 \label{Exc_dis_study}
\end{figure}

In figure \ref{Exc_dis_study} we show the effective temperature of the excitons at a lattice temperature of $T_{lattice}=$\unit[7]{K} for different exciton lifetimes $\tau_{dark}$. To obtain more insights, we also show the effective exciton temperature after pulsed excitation. First we find a general increase of the effective temperature as the lifetime decreases: excitons with short lifetimes do not have
enough time to thermalize before they recombine (dark), and therefore the exciton distribution stays ”hot”. If only radiative recombination is taken into account, we find that the effective temperature is only slightly larger compared to the pulsed excitation. Interestingly the effective temperature after pulsed excitation is \unit[9]{K}, being \unit[2]{K} larger than the phonon temperature. The reason for this behavior is the radiative decay of excitons within the lightcone, which disturbes the exciton distribution. We have also performed calculations with artificially turned off radiative decay, and the effective temperature was \unit[7.06]{K} where the residual deviation from the lattice temperature of about 1\% stems from numerical uncertanties (the finite grid resolution enters as a lower bound for how cold the excitons can be).

We find that the effective exciton temperature increases monotonously with decreasing lifetime, where all curves show a qualitatively similar behavior. First we find an overall increasing effective temperature as a function of detuning, since for small detunings only acoustic phonon scattering contributes to the thermalization, which is slow at \unit[7]{K}. However, this increasing trend is interupted at \unit[30]{meV} and \unit[36]{meV}, since for this excitation energy, the generated excitons can swiftly relax via the emission of optical phonons, leading to a cooling of the excitons effectively.

\textit{Conclusion:} In conclusion we have presented microscopic calculations for the non-equilibrium steady state in monolayer MoSe$_2$ during continuous wave laser excitation. In agreement with available experimental data, we have revealed that phonon mediated relaxation via emission of one optical $\Gamma$ phonon or two optical and acoustic zone-edge phonons causes the appearance of pronounced phonon bands in the PLE. Further the DoP as a function of the laser detuning exhibits similar sidebands due to the domination of phonon assisted relaxation over intervalley exchange coupling. Our results for the exciton momentum distribution of excitons indicate, that during cw pumping significant heating of excitons up to \unit[40]{K} can be achieved.

\textit{Acknowledgement:} We acknowledge great discussion with Hans Tonatzky (CNRS Toulouse) and Daniel Wigger (Wrocław University of Science and Technology). This work was funded by the Deutsche Forschungsgemeinschaft (DFG) - Projektnummer 182087777 - SFB 951 (project B12, M.S.,D.C.,M.K., A.K.). M.S. gratefully acknowledges funding from the Deutsche Forschungsgemeinschaft through Project No. 432266622 (SE 3098/1). This work was supported by HFML-RU/NWO-I , member of the 
European Magnetic Field Laboratory (EMFL).

\end{document}